\documentclass[12pt]{iopart}
\usepackage{graphicx}
\usepackage{wrapfig}

\begin{document}

\title{$\Upsilon$ cross section in p+p collisions at STAR}

\author{Pibero Djawotho for the STAR Collaboration}

\address{Indiana University Cyclotron Facility, Bloomington, IN 47408, USA}
\ead{pibero@indiana.edu}
\begin{abstract}
The main focus of the heavy flavor program at RHIC is to
investigate the properties of the dense matter produced
in heavy-ion collisions by studying its effect on open heavy
flavor and quarkonia production. This in turn requires a detailed
understanding of their production in elementary p+p collisions
so that the dense matter effects can be later unfolded.
In this paper, we present the first mid-rapidity cross section
measurement of bottomonium at $\sqrt{s}=200$ GeV with the STAR
experiment. We compare our results with perturbative QCD
calculations. A brief status on the study of charmonium in STAR
is given.
\end{abstract}


\section{Introduction}
$J/\psi$ suppression induced by Debye screening of the static potential between $c\bar{c}$
pairs was originally hailed as an unambiguous signature of the QGP~\cite{matsui}.
This simple picture, however, is complicated by competing effects
that either reduce the yield, such as comover absorption~\cite{gavin, blaizot},
or enhance it, such as recombination models~\cite{grandchamp, thews, andronic}.
With the low cross section of the $\Upsilon$ family, the roles of absorption and
recombination are negligible.
Lattice QCD studies of quarkonia spectral functions indicate that while
the $\Upsilon''$ melts at RHIC and the $\Upsilon'$ is likely to melt,
the $\Upsilon$ is expected to survive~\cite{digal, wong}.
Suppression of the $\Upsilon'$ and $\Upsilon''$ should be measurable at RHIC energies
over a wide range of $p_T$. Therefore, it is only with a systematic study of all quarkonia
states in pp, pA and AA systems that a clear understanding of the properties of dense matter
will emerge~\cite{rhic2}. In this paper, we report preliminary results of the $\Upsilon$
cross section at midrapidity obtained with the STAR detector in p+p collisions at
$\sqrt{s}=200$ GeV via the dielectron decay channel.

\section{Detector Overview}
The main detectors used in the STAR quarkonia program are the TPC
(Time Projection Chamber)~\cite{tpc} and the EMC (Electromagnetic Calorimeter)~\cite{emc}.
They both have a large acceptance: pseudorapidity $|\eta|<1$ and full azimuthal coverage.
The combined capabilities of the TPC+EMC for electron identification,
together with the triggering capabilities of the EMC are the two pillars
of the STAR quarkonium program. In particular, the EMC trigger allows
us to sample the full luminosity delivered by RHIC to look for the
high-mass di-electron signals characteristic of the $\Upsilon\rightarrow e^+e^-$ decay.
We now summarize the quarkonia triggers developed for this purpose.

\section{The STAR Quarkonia Triggers}
The architecture of the STAR quarkonia trigger is based on a two-level system comprising
a level-0 (L0) hardware component ($\sim 1~\mu$s) and a level-2 (L2) software component
($\sim 100~\mu$s).
L0 requires an EMC tower to have energy above 3.5 GeV and the associated trigger patch
consisting of $4\times 4$ towers to have total energy above 4.3 GeV, coupled with
a minimum bias condition.
Computer simulations show that $\Upsilon$ with both daughter electrons having energy above
3 GeV account for 95\% of all $\Upsilon$ that decay within the acceptance of the EMC.
The advantage of triggering at such high-energy is the added hadron rejection power
$e/h\sim 100$ of the EMC towers. L2 performs tower clustering to reclaim energy leaked into
neighboring towers and improve mass resolution. Cuts are applied on the invariant mass
$m_{ee}=\sqrt{2E_1E_2(1-\cos\theta_{12})}$ and the opening angle $\theta_{12}$ between clusters.
The relatively clean environment of the higher mass states permits the trigger to be
utilized in Au+Au collisions as well. The trigger is highly efficient and limited only by
luminosity affording the STAR detector with its large acceptance a position of prominence for
$\Upsilon$ measurements. Because of the small $\Upsilon$ cross section, large luminosity and
acceptance are necessary. The $J/\psi$ L0-L2 trigger chain is very similar to that of the
$\Upsilon$. It is suitable only in p+p and d+Au collisions since in high multiplicity
nuclear collisions the trigger rejection drops dramatically. The trigger was successfully
tested in 2005 and 2006, and the $J/\psi$ analysis is in progress.

\section{$\Upsilon$ Analysis and Result}
\begin{figure}
\begin{minipage}[b]{.46\linewidth}
\includegraphics*[width=7cm]{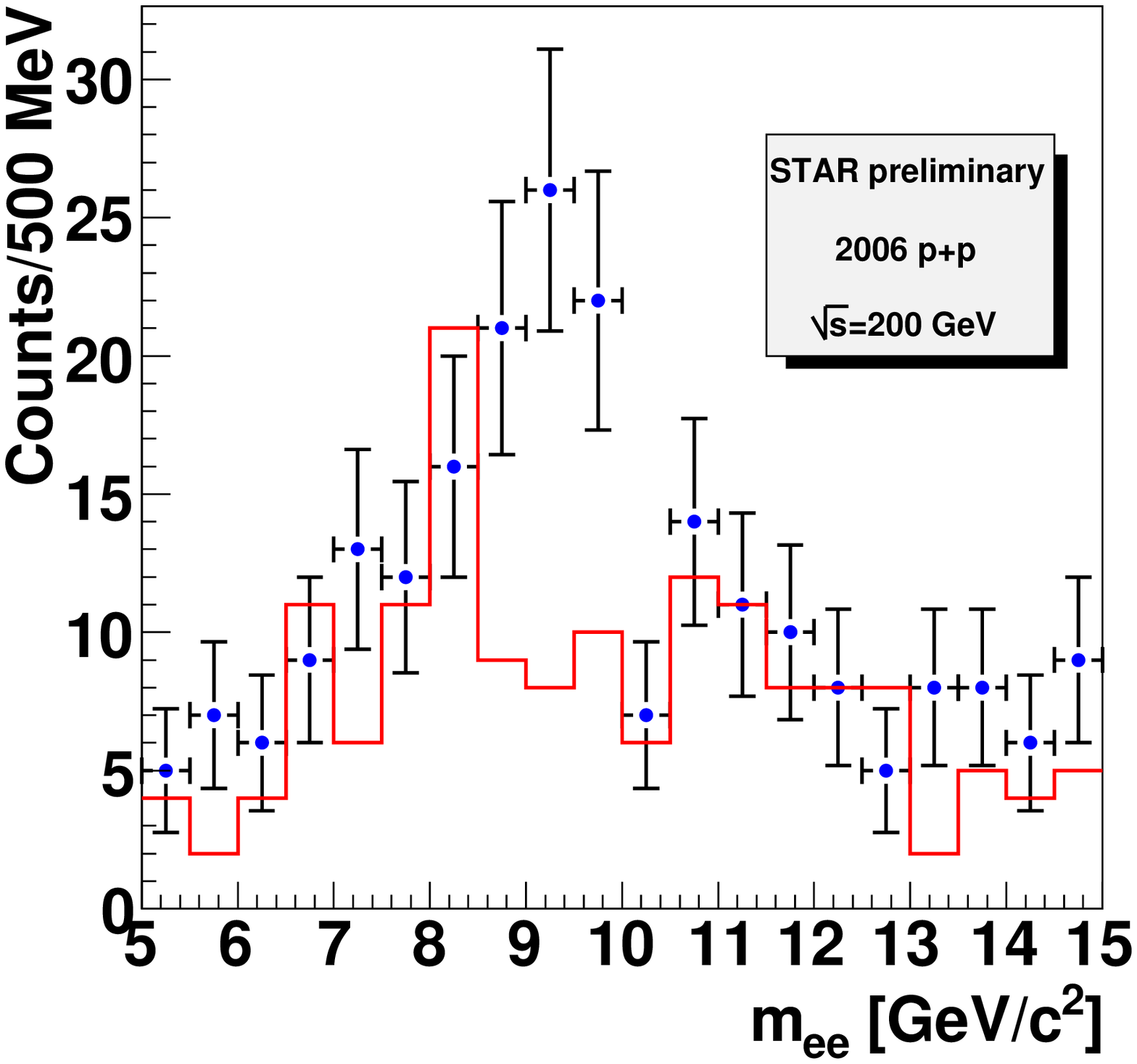}
\end{minipage}\hfill
\vspace{-2mm}
\begin{minipage}[b]{.46\linewidth}
\includegraphics*[width=7cm]{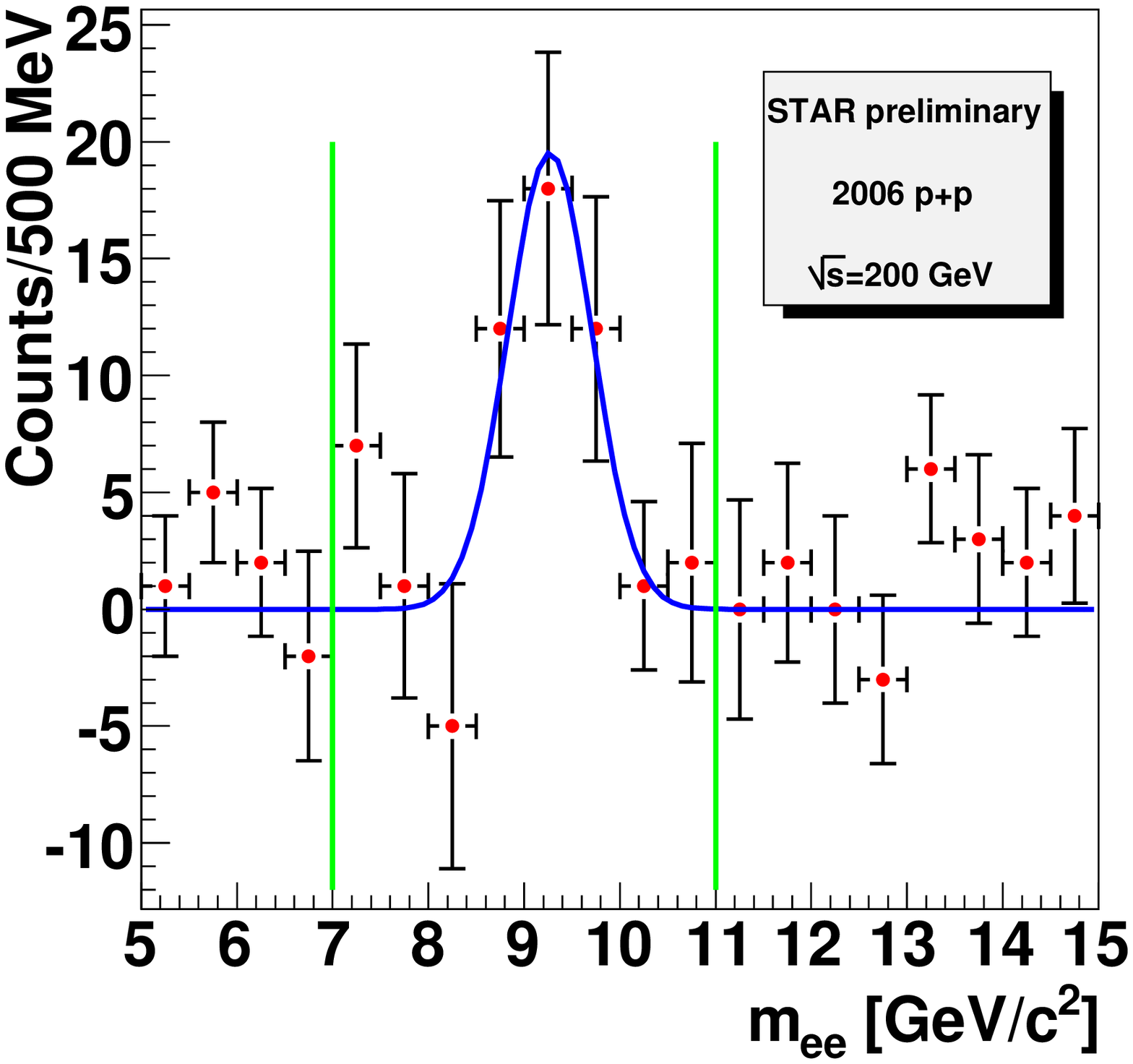}
\end{minipage}
\vspace{-2mm}
\caption{Left panel: STAR 2006 p+p collisions $\Upsilon \rightarrow e^+e^-$
at $\sqrt{s}=200$ GeV signal and background  with statistical error bars
from combining unlike-sign pairs (electrons and positrons). The red background results
from combining like-sign pairs. Right panel: Background-subtracted $\Upsilon$ signal with
statistical error bars. The blue curve is a Gaussian fit. The green vertical bars mark
the boundaries of integration for the yield.}
\label{fig:upsilon_peak}
\end{figure}

In 2006, with full EMC acceptance, STAR sampled 9 pb$^{-1}$ of integrated luminosity. 
Two different trigger setups were deployed but the analysis focused on one setup with
integrated luminosity $\int\mathcal{L}dt=(5.6\pm0.8)~{\rm pb}^{-1}$ (syst).
Electrons were identified by selecting charged particle tracks with specific $dE/dx$
ionization energy loss in the TPC that deposited more than 3 GeV of energy in an EMC tower.
Electron-positron pairs were then combined to produce the invariant mass spectrum.
Finally, like-sign electron pairs were combined to form the invariant mass spectrum of
the background which was subtracted from the signal and background;
see Fig.~\ref{fig:upsilon_peak}.

\begin{figure}
\begin{minipage}[b]{.46\linewidth}
\includegraphics*[height=8.5cm]{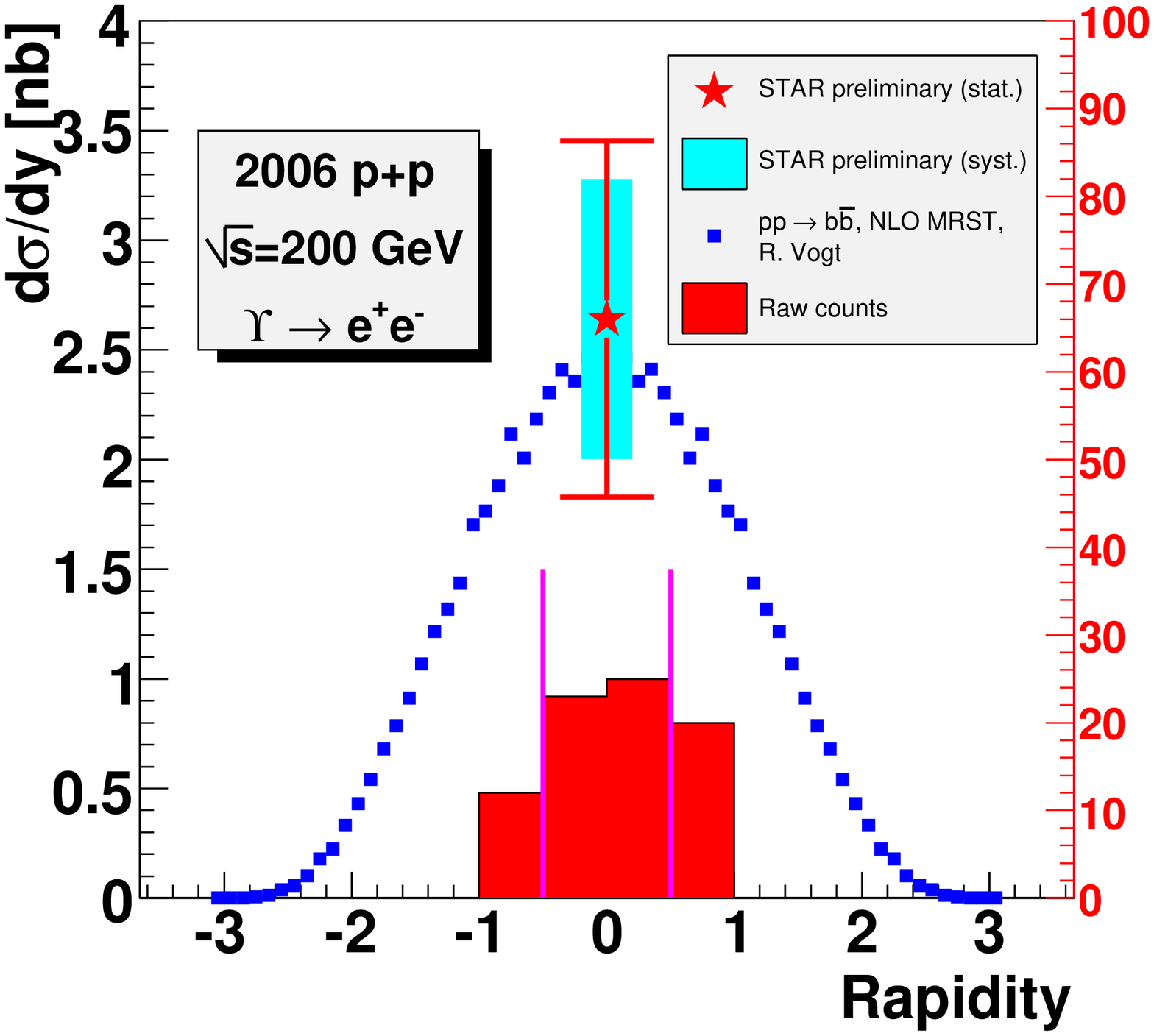}
\end{minipage}\hfill
\vspace{-2mm}
\begin{minipage}[b]{.40\linewidth}
\includegraphics*[height=7.5cm]{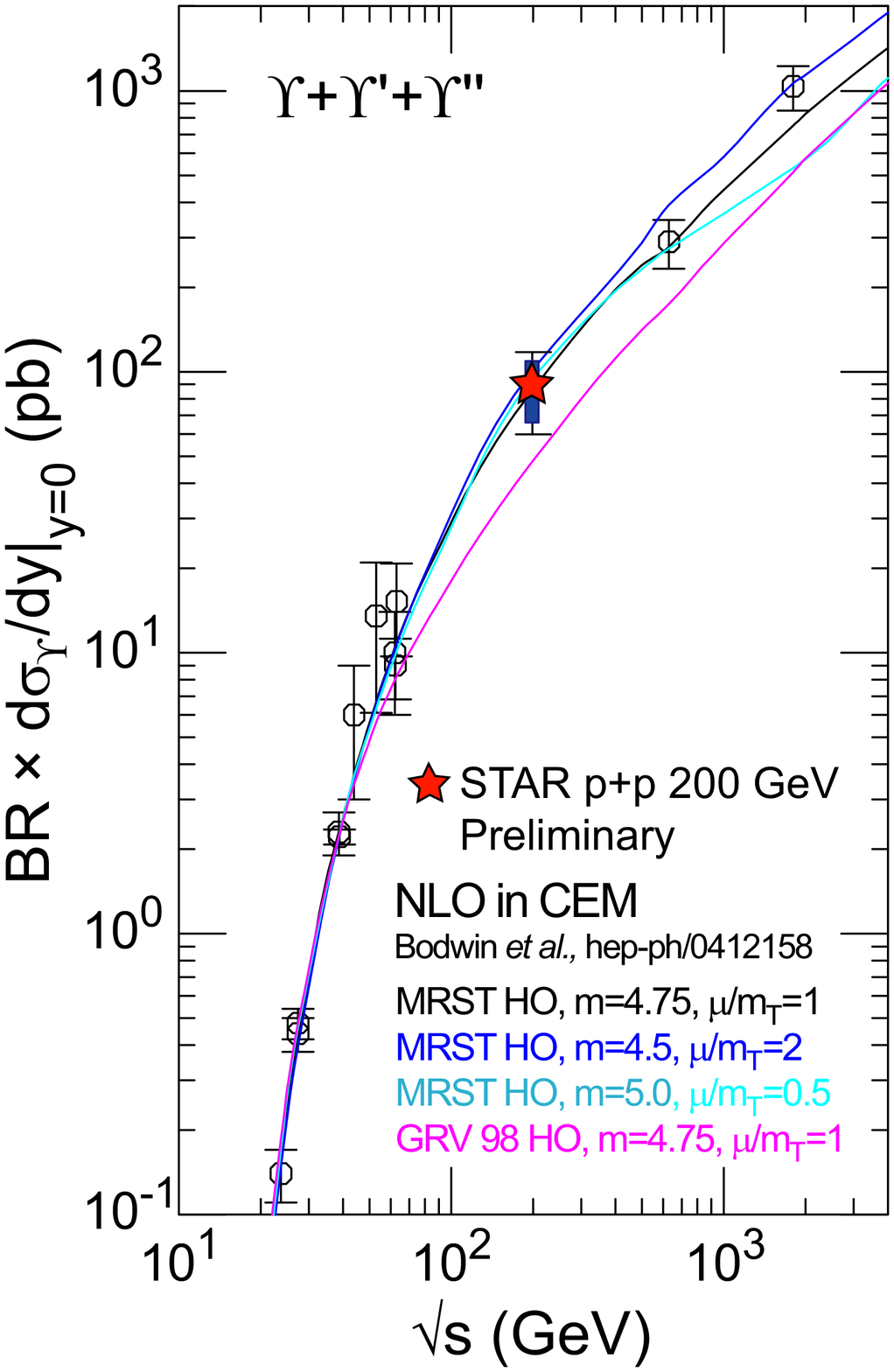}
\end{minipage}
\vspace{-2mm}
\caption{Left panel: STAR $\Upsilon \rightarrow e^+e^-$ differential cross section 
on the left axis {\it vs.} rapidity compared with NLO pQCD calculations~\cite{vogt}.
The red middle histogram shows the raw counts with axis on the right. The magenta vertical
bars identify the limits of integration $|y|<0.5$.
Right panel: $\Upsilon+\Upsilon'+\Upsilon''$ cross section at midrapidity {\it vs.}
center-of-mass energy. The curves are different NLO CEM predictions~\cite{rhic2}}
\label{fig:upsilon_cross_section}
\end{figure}

STAR simulation of the bottomonia states show that the STAR detector cannot resolve
individual states of the $\Upsilon$ family due to finite momentum resolution
and electron bremsstrahlung given the current material budget between the interaction vertex
and the TPC entrance window. Therefore, the yield reported here is for the combined
$\Upsilon+\Upsilon'+\Upsilon''$ states. The total yield was extracted by integrating
the invariant mass spectrum from 7 to 11 GeV/$c^2$. The width of the peak $\sim$ 1 GeV/$c^2$
was found to be consistent with simulation. The significance of the signal was estimated at
$3\sigma$. The cross section was calculated with the formula:

\begin{eqnarray}
\sum_{i=\Upsilon,\Upsilon',\Upsilon''}BR_i\times\left(\frac{d\sigma}{dy}\right)^i_{y=0}=
\frac{N}{dy\times\epsilon\times\int\mathcal{L}dt},
\label{eq:upsilon}
\end{eqnarray}

where $BR_i$ is the branching ratio for
$\Upsilon,\Upsilon',\Upsilon''\rightarrow e^+e^-$~\cite{pdg},
$N=48\pm 15$ (stat.) is the raw yield, $dy=1.0$ is the rapidity interval and
$\epsilon=\epsilon_{\rm geo}\times\epsilon_{\rm L0}\times\epsilon_{\rm L2}\times\epsilon^2(e)
\times\epsilon_{\rm mass}=(9.4\pm 1.8)\%$ (syst.) is the efficiency for reconstructing
members of the $\Upsilon$ family. $\epsilon_{\rm geo}$ is the geometrical acceptance,
$\epsilon_{\rm L0}$ and $\epsilon_{\rm L2}$ are the trigger efficiencies for L0 and L2,
respectively. $\epsilon(e)$ is the efficiency for reconstructing a single electron and
$\epsilon_{\rm mass}$ is the efficiency of the invariant mass cut.
We find for the cross section at midrapidity in $\sqrt{s}=200$ GeV p+p collisions
$BR\times(d\sigma/dy)^{\Upsilon+\Upsilon'+\Upsilon''}_{y=0}=91\pm 28~{\rm (stat.)}
\pm 22~{\rm (syst.)~pb}$. The systematic error is dominated by the uncertainty in the
integrated luminosity.
Fig.~\ref{fig:upsilon_cross_section} (left panel) shows the STAR $\Upsilon$ cross section
{\it vs.} rapidity along with NLO pQCD calculations~\cite{vogt}.
Fig.~\ref{fig:upsilon_cross_section} (right panel) shows the STAR $\Upsilon$
measurement along with the world data and pQCD predictions {\it vs.} center-of-mass energy.

\section{Conclusion}
The STAR experiment made the first RHIC measurement of the
$\Upsilon,\Upsilon',\Upsilon''\rightarrow e^+e^-$ cross section at midrapidity in p+p collisions
at $\sqrt{s}=200$ GeV; $BR\times(d\sigma/dy)_{y=0}=91\pm28~{\rm (stat.)}\pm22~{\rm (syst.)}$
The STAR $\Upsilon$ measurement is consistent with the world data and NLO in the CEM
(Color Evaporation Model) pQCD calculations~\cite{rhic2}.
The full EMC and a trigger are essential for a successful quarkonia program at STAR.
With full EMC coverage, STAR looks to make a significantly improved measurement
of the $\Upsilon$ cross section in Au+Au over the 2004 upper limit on the $\Upsilon$
cross section with only half the EMC~\cite{kollegger}.
It is projected that with RHIC I luminosity and detector upgrades, STAR will sample
an estimated yield of 6,600 $\Upsilon$ in p+p and 830 $\Upsilon$ in Au+Au.
RHIC II luminosity should boost the $\Upsilon$ yield to 8,300 in p+p and 11,200 in
Au+Au~\cite{rhic2}.
The author acknowledges the U.S. National Science Foundation's support under
grant number NSF-PHY-0457219.

\end{document}